\documentclass[prb,letterpaper,aps,floatfix,twocolumn]{revtex4-2}

\usepackage{graphicx}
\usepackage{amsmath,amssymb}
\usepackage{bm}

\usepackage[pdfstartview=FitH,breaklinks=true,bookmarks=true,colorlinks=true,anchorcolor=black,citecolor=red,filecolor=black,menucolor=black,urlcolor=blue,linkcolor=blue]{hyperref}

\usepackage{color}
\def\Red{\color{black}}

\begin{document}

\title{Influence of flat bands on RKKY interaction: perspective of Fano defects}

\author{Yue-De Luo}
\affiliation{Department of Applied Physics, Tunghai University, Taichung 40704, Taiwan}
\author{Min-Fong Yang}
\email{mfyang@thu.edu.tw}
\affiliation{Department of Applied Physics, Tunghai University, Taichung 40704, Taiwan}

\date{\today}

\begin{abstract}
In this paper, we revisit the effect of flat bands on the Ruderman-Kittel-Kasuya-Yosida (RKKY) interaction by using a coordinate transformation that detangles flat-band states from dispersive ones. Under this transformation, original flat-band systems containing magnetic impurities are mapped onto a generalized Fano-Anderson model, where flat-band states act as Fano defects. From this perspective, several features of exact RKKY couplings calculated numerically can be understood easily. As an illustrative example, we analyze a dimerized diamond chain model, which can exhibit either gapped or gapless spectra depending on the ratio of hopping integrals. We find that anomalous decay in the RKKY couplings arises exclusively in the gapless case and with specific magnetic coupling configurations. Furthermore, the conventional wisdom regarding the signs of RKKY interactions breaks down under certain conditions. These subtleties arising from flat bands find explanation within our present approach.
Our investigation offers deeper insights into how flat bands influence carrier-mediated exchange interactions, with implications extending to broader contexts.
\end{abstract}

\maketitle

\section{INTRODUCTION} \label{sec:intro}

Systems hosting flat bands have captured significant interest in condensed matter physics over the past decades. Various theoretical models and artificially constructed flat-band materials have been founded and extensively studied~\cite{FB-Review-Flach,FB-Review-Chen,FB-Review-Vicencio,%
FB-classification-Calugaru,FB-catalog-npj-Comp-Mat,Danieli_etal2024}. In these systems, the flat energy bands spanning over the entire Brillouin zone arise from destructive interference of hoppings between neighboring sites. Eigenstates corresponding to these flat-band energies can be described by the compact localized states (CLSs)~\cite{Sutherland1986,Flach_etal2014,Bodyfelt_etal2014,Danieli_etal2015}, whose amplitudes are nonzero only within a region spanning a few unit cells.
%
%
Due to quenching of kinetic energy and macroscopic degeneracy, flat-band systems are sensitive to perturbations, allowing intriguing correlated states to emerge when interactions are at play, such as ferromagnetism~\cite{Tasaki-Review,Derzhko_etal-Review}, superconductivity~\cite{Aoki-Review}, and fractional topological insulators~\cite{Bergholtz-Review}.

In recent years, there has been significant interest in understanding the effects of flat bands on the Ruderman-Kittel-Kasuya-Yosida (RKKY) interaction~\cite{Ruderman-Kittel1954,Kasuya1956,Yosida1957}. The RKKY interaction is an indirect exchange interaction between two localized magnetic moments mediated by a background of electrons. In the context of zigzag graphene nanoribbons, it was found that the conventional non-degenerate second-order perturbation theory falls short in explaining all aspects of the RKKY interaction due to the presence of flat bands~\cite{Bunder2009,Black-Schaffer2010,Cao2019}. The same conclusion is reached as well when studying two-dimensional flat-band lattice models~\cite{Oriekhov2020,Bouzerar2021,Bouzerar2022}. This may not be surprising, since the vanishing bandwidth and the high degeneracy of a flat band can render the standard RKKY approximation invalid.

To explore the applicability and limitations of perturbation theory, K. Laubscher~\emph{et al.} study the RKKY interaction in two one-dimensional flat-band models at half filling~\cite{Laubscher_etal2023}: the stub lattice (with a gapped spectrum) and the diamond lattice (with a gapless spectrum). Their numerically exact calculations reveal peculiar features in the RKKY interaction that defy the conventional non-degenerate perturbation theory.
For conventional one-dimensional systems with a partially filled band, the RKKY coupling decays as $1/R$ ($R$ is the inter-impurity distance). However, they find that the RKKY coupling in the stub lattice decays exponentially with $R$ even though the flat band is partially filled. For the case of the diamond lattice, the RKKY coupling can show unusual asymptotic decay of either $1/R^3$ or $1/R^5$ depending on the sublattice configuration.
They conclude that, when an energy gap exists between the flat and the dispersive bands such as the case of the stub lattice, degenerate perturbation theory yields good agreement with numerical data. However, for systems with a gapless spectrum like the diamond lattice, a nonperturbative approach becomes essential.

In this paper, we pursuit this issue to get further understanding of the RKKY interaction in flat-band systems. We consider here a dimerized diamond chain, in which the hopping integrals within and between unit cells can be different, as shown in the left of Fig.~\ref{fig:dimond_chain}. The considered model has two dispersive and one flat bands. By adjusting the ratio $\lambda$ of the two hoppings, this flat-band model can encompass either a gapped or gapless spectrum. We examine the generalized case in which magnetic impurities resided on the $B$ sublattice can couple to both lattice sites $B$ and $C$ with a strength ratio of $x$.
As highlighted in  Refs.~\cite{Flach_etal2014,Flach_etal2014,Bodyfelt_etal2014,Danieli_etal2015}, the role played by flat-band states can be made clearer when employing a coordinate transformation to detangle them from the dispersive states. Under this procedure, our system is mapped onto a generalized Fano-Anderson model, in which the coupling to magnetic impurities results in a hybridization of localized flat-band states with a linear chain (see the right of Fig.~\ref{fig:dimond_chain}). It is shown below that the degree of hybridization between the flat and the dispersive bands can be tuned by varying $x$.

By analyzing the detangled model, we find that the flat band is usually inconsequential, and the conventional result is restored up to a proportional factor. Anomalous decay laws in the RKKY couplings observed in Ref.~\cite{Laubscher_etal2023} arise exclusively in the gapless case (where the hopping ratio $\lambda=1$) and with a magnetic coupling ratio of $x=0$. Interestingly, when $\lambda=1$ and $x<0$, the sign of RKKY coupling between different sublattices does not follow the general result for bipartite lattices at half filling~\cite{Saremi2007}, even though the conventional $1/R$ decay is obeyed. All these features can be well understood within the present approach.
Although we focus on a specific one-dimensional model, the conclusions of our study would be applicable to broader contexts.

The remainder of this paper is organized as follows.
We describe our model in Sec.~\ref{sec:model}. We then introduce the coordinate transformation and derive the corresponding detangled form of our model.
The results for both the gapped ($\lambda\neq1$) and the gapless ($\lambda=1$) cases are presented in Sec.~\ref{sec:results}.
We conclude our paper in Sec.~\ref{sec:conclusion}.
{\Red %
In the Appendix~\ref{sec:App1}, dependence of the RKKY couplings on the Kondo coupling strengths is discussed. An effective on-site potential employed in Sec.~\ref{sec:lambda1} is derived in Appendix~\ref{sec:App2}.
}%

\section{dimerized diamond chain and its detangled form}\label{sec:model}

\begin{figure}[htb]
\includegraphics[width=0.95\linewidth]{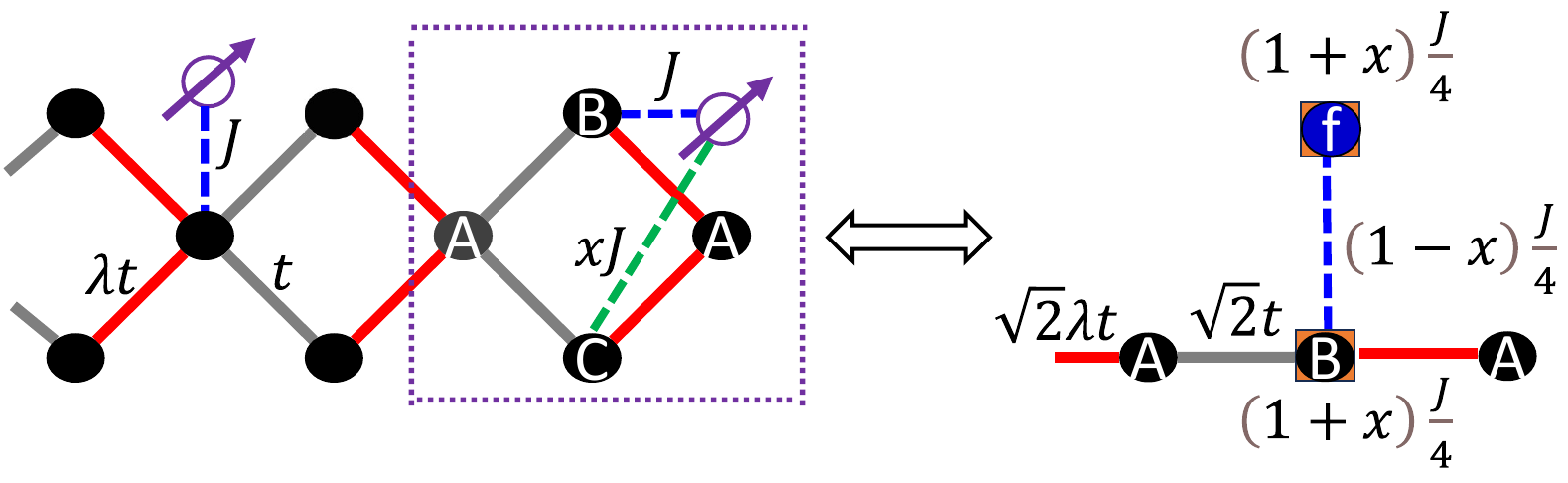}
\caption{The dimerized diamond lattice (left) and its detangled form (right). The intra-unitcell (gray) and the inter-unitcell (red) hoppings have different strengths $t$ and $\lambda t$, respectively. In the detangled representation, the coupling of strength $(1-x)J/4$ between the localized flat-band state $f_n$ and the linear chain is induced by the magnetic coupling $J$ of the magnetic moment. In addition, there appear effective on-site impurity potentials of strength $(1+x)J/4$.}
\label{fig:dimond_chain}
\end{figure}

In the present work, we consider a dimerized diamond chain, whose band structure contains a flat band and can be either gapless or gapped by tuning the hopping ratio. As shown in the left of Fig.~\ref{fig:dimond_chain}, the unit cell consists of three sites, say $A$, $B$, and $C$. Just like the undimerized case, this lattice is a bipartite one with one sublattice composed of all $A$ sites and the other containing all $B$ and $C$ sites. The Hamiltonian is given by
\begin{align}
H_0=&\sum_{n,\sigma}\Big[ t \left( c_{n,A,\sigma}^\dagger c_{n,B,\sigma} + c_{n,A,\sigma}^\dagger c_{n,C,\sigma} \right) \nonumber\\
& +\lambda t \left( c_{n+1,A,\sigma}^\dagger c_{n,B,\sigma} + c_{n+1,A,\sigma}^\dagger c_{n,C,\sigma} \right) \Big] + \mathrm{H.c.} \; ,
\label{eq:Hdiamond}
\end{align}
where the intra-unitcell and the inter-unitcell hoppings are denoted by $t$ and $\lambda t$, respectively. $\sigma=\pm$ denotes the spin orientation.
The bulk spectrum consists of a zero-energy flat band $\epsilon_0(k)=0$ and two dispersive bands $\epsilon_\pm(k)=\pm\sqrt{2}t\sqrt{(1-\lambda)^2+4\lambda\cos^2{(k/2)}}$. Here the size of unit cells is set to unity. When $\lambda\neq1$, there is an energy gap $\Delta=\sqrt{2}|1-\lambda|t$ at $k=\pi$ separating the flat band from the dispersive bands. The flat band can be described in terms of a set of CLSs having support on two lattice sites within a single unit cell each. The creation operators corresponding to these eigenmodes are given by $f_{n,\sigma}^\dagger=(c_{n,B,\sigma}^\dagger-c_{n,C,\sigma}^\dagger)/\sqrt{2}$.
Throughout this work, we set the Fermi level $\epsilon_F=0$ and focus on the case of a half-filled flat band. Actually, as mentioned in Ref.~\cite{Laubscher_etal2023}, the results do not depend on the exact filling factor as long as the flat band stays partially filled.

We then place two classical impurity spins $\mathbf{S}_1$ and $\mathbf{S}_2$ in the unit cells $n_1$ and $n_2$ at sublattice positions $\alpha$ and $\beta$ ($\alpha,\beta=A$, $B$), respectively. For simplicity, we normalize the impurity spins such that $|\mathbf{S}_i|=1$. As usual, the interaction between the impurity spins and the itinerant electrons is assumed to be a simple Kondo coupling term. When the impurity spin $\mathbf{S}_i$ locates at $A$ sublattice, the local exchange coupling reads
\begin{equation}\label{eq:impurtyA}
H_\mathrm{imp}^{(i)}=\frac{J_i}{2}\sum_{\sigma,\sigma'}\, c_{n_i,A,\sigma}^\dagger\, [\mathbf{S}_i \cdot\boldsymbol{\sigma}]^{\sigma\sigma'} c_{n_i,A,\sigma'} \; .
\end{equation}
Here $\boldsymbol{\sigma}$ is the vector of Pauli matrices and the spin labels are denoted by $\sigma,\sigma'=\pm$.
On the other hand, when the impurity spin $\mathbf{S}_i$ locates at $B$ sublattice and it couples as well to site $C$ of the same unit cell with a relative strength $x$, the magnetic coupling then becomes
\begin{align}
H_\mathrm{imp}^{(i)}=&\frac{J_i}{2}\sum_{\sigma,\sigma'}\, c_{n_i,B,\sigma}^\dagger\, [\mathbf{S}_i \cdot\boldsymbol{\sigma}]^{\sigma\sigma'} c_{n_i,B,\sigma'} \nonumber\\
&+\frac{xJ_i}{2}\sum_{\sigma,\sigma'}\, c_{n_i,C,\sigma}^\dagger\, [\mathbf{S}_i \cdot\boldsymbol{\sigma}]^{\sigma\sigma'} c_{n_i,C,\sigma'} \; .
\label{eq:impurtyB}
\end{align}
Without loss of generality, we have $|x|\leq 1$.

Because our model has spin rotational symmetry, the carrier-mediated indirect exchange interaction between these two impurity spins will take an effective Heisenberg form,
\begin{equation}\label{eq:H_RKKY}
H_\mathrm{RKKY}=J_\mathrm{RKKY}^{\alpha\beta}(R)\;\mathbf{S}_1\cdot\mathbf{S}_2 \; .
\end{equation}
The effective RKKY coupling constant $J_\mathrm{RKKY}^{\alpha\beta}(R)$ depends on the sublattice positions $\alpha$ and $\beta$ ($\alpha,\beta=A$, $B$) of the impurities and on the inter-impurity distance $R=n_2-n_1>0$.

We note that, because the total Hamiltonian $H_0+H_\mathrm{imp}^{(1)}+H_\mathrm{imp}^{(2)}$ is quadratic, the accurate ground-state energies for arbitrary impurity spin orientations $\{\mathbf{S}_1,\mathbf{S}_2\}$ and separation distances $R$ can be evaluated numerically via exact diagonalization. According to Eq.~\eqref{eq:H_RKKY}, the exact RKKY coupling can be determined by~\cite{Black-Schaffer2010,Deaven_etal1991}
\begin{equation}\label{eq:RKKY_energy_difference}
J_{\mathrm{RKKY}}^{\alpha\beta}= (E_{\mathrm{FM}}^{\alpha\beta}-E_{\mathrm{AFM}}^{\alpha\beta})/2 \; .
\end{equation}
Here $E_{\mathrm{FM}}^{\alpha\beta}$ and $E_{\mathrm{AFM}}^{\alpha\beta}$ correspond to the exact ground state energies for the ferromagnetic (FM) and antiferromagnetic (AFM) configuration of spin impurities. These energies are computed with respect to an arbitrarily chosen spin quantization axis, such as the $z$ axis, where $\mathbf{S}_i=S_i\mathbf{e}_z$ with $S_i=\pm1$.

To analyze the numerical data, we follow the detangle procedure proposed in Refs.~\cite{Flach_etal2014,Flach_etal2014,Bodyfelt_etal2014,Danieli_etal2015}. For the considered model, it can be achieved by the coordinate transformations local to the unit cells and defined by the real matrix $\hat{U}$:
\begin{equation}
\left(
\begin{array}{ccc}
d_{n,A,\sigma} \\
d_{n,B,\sigma} \\
f_{n,\sigma}
\end{array} \right) = \hat{U} \; \left(
\begin{array}{ccc}
c_{n,A,\sigma} \\
c_{n,B,\sigma} \\
c_{n,C,\sigma}
\end{array} \right) , \quad
\hat{U} = \frac{1}{\sqrt{2}}\left(
\begin{array}{ccc}
\sqrt{2} & 0 & 0 \\
0 & 1 & 1 \\
0 & 1 & -1
\end{array} \right)\; .
\label{eq:fano-DC}
\end{equation}
In terms of new variables,
\begin{align}
\tilde{H}_0=\sum_{n,\sigma} &\left( \sqrt{2}t\;d_{n,A,\sigma}^\dagger d_{n,B,\sigma} \right. \nonumber \\
&\left. + \sqrt{2}\lambda t\; d_{n+1,A,\sigma}^\dagger d_{n,B,\sigma} \right) + \mathrm{H.c.} \; ,
\label{eq:detangled0}
\end{align}
such that the CLSs described by $f_{n,\sigma}$ are completely detangled from the dispersive states $d_{n,A,\sigma}$ and $d_{n,B,\sigma}$ of a single chain. Notice that the CLSs appear only on top of $B$ sublattice (see the right of Fig.~\ref{fig:dimond_chain}). At half filling, the lowest dispersive band with energy $\epsilon_-(k)$ is completely occupied, while the zero-energy flat band is half filled.
In the presence of an impurity spin $\mathbf{S}_i=S_i\mathbf{e}_z$ located at $B$ sublattice and couple to site $C$ simultaneously, the local exchange coupling in Eq.~\eqref{eq:impurtyB} is transformed to
\begin{align}
\tilde{H}_\mathrm{imp}^{(i)}=&\sum_{\sigma}\,V^{(i)}_\sigma\left( d_{n_i,B,\sigma}^\dagger\,d_{n_i,B,\sigma} + f_{n_i,\sigma}^\dagger\,f_{n_i,\sigma} \right) \nonumber\\
&+\sum_{\sigma}\,t^{(i)}_\sigma\left( d_{n_i,B,\sigma}^\dagger\,f_{n_i,\sigma} + \mathrm{H.c.} \right)
\label{eq:detangledB}
\end{align}
with
\begin{align}
&V^{(i)}_\sigma=\frac{1}{4}\sigma S_i(1+x)J_i \; , \label{eq:V} \\
&t^{(i)}_\sigma=\frac{1}{4}\sigma S_i(1-x)J_i \; . \label{eq:t}
\end{align}
Here $V^{(i)}_\sigma$ represents a spin-dependent on-site energy and $t^{(i)}_\sigma$ gives a spin-dependent hybridization between the CLS and the dispersive state, as shown in the right of Fig.~\ref{fig:dimond_chain}.
On the other hand, if the impurity spin is located at $A$ sublattice, on which the flat-band states do not have support, the transformed local exchange coupling for Eq.~\eqref{eq:impurtyA} becomes
\begin{equation}\label{eq:detangledA}
\tilde{H}_\mathrm{imp}^{(i)}=\sum_{\sigma}\,V^{(i)}_\sigma\, d_{n_i,A,\sigma}^\dagger\,d_{n_i,A,\sigma}
\end{equation}
with $V^{(i)}_\sigma=\sigma S_iJ_i/2$.

The detangled Hamiltonian displayed above is nothing but a generalized Fano-Anderson model with local on-site energies and hybridizations induced by impurities, where the CLSs act as Fano defects~\cite{Flach_etal2014,Flach_etal2014,Bodyfelt_etal2014,Danieli_etal2015}. From this representation, it becomes evident that only the CLSs situated atop the impurity sites play a significant role, while other states do not contribute. Consequently, whether the latter states are occupied by electrons or not becomes irrelevant. This observation readily explains why the results reported in Ref.~\cite{Laubscher_etal2023} do not depend on the precise filling factor, as long as the flat band remains partially filled.

Furthermore, when magnetic impurities couple both lattice sites $B$ and $C$ with equal strength (i.\,e., $x=1$), $t^{(i)}_\sigma=0$, resulting in complete detangling of the CLSs. In this case, the conventional one-dimensional result is expected. This conclusion is verified by the following numerical calculations.

\section{results}\label{sec:results}

In this section, we present our numerical data of the exact RKKY couplings in the dimerized diamond lattice for various $\lambda$'s and $x$'s by using Eq.~\eqref{eq:RKKY_energy_difference}. These results can be well understood by using the detangled model in Eqs.~\eqref{eq:detangled0}, \eqref{eq:detangledB}, and~\eqref{eq:detangledA}. Without loss of generality, we set $t=1$ as the energy unit and $J_1=J_2=0.2\,t$.

\subsection{Gapped systems with $\lambda\neq1$}\label{sec:lambda2}

We start by exploring the RKKY interaction for the cases of $\lambda\neq1$, where an energy gap exists between the flat and the dispersive bands. For simplicity, we set $x=0$ such that there is no local exchange coupling to the $C$ sublattice.

\begin{figure}[htb]
\includegraphics[width=0.9\linewidth]{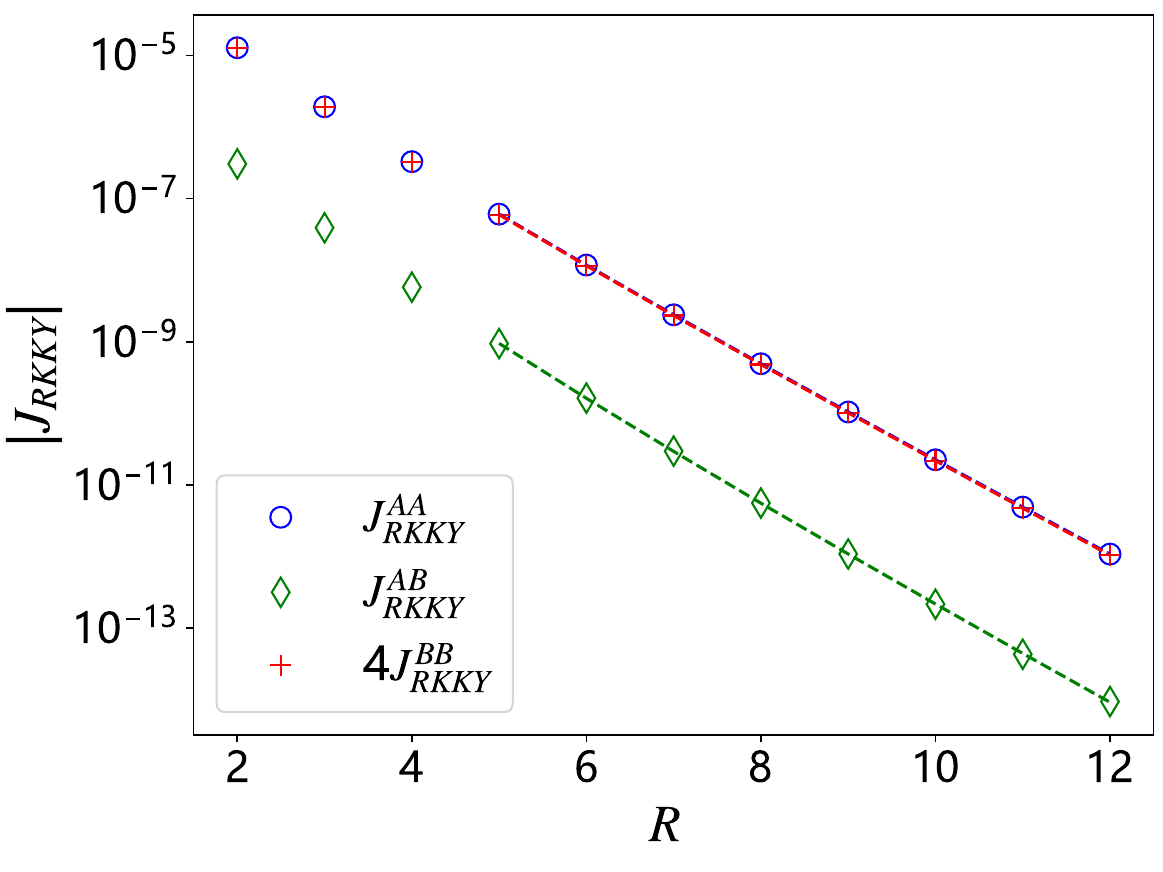}
\caption{Absolute value of the RKKY couplings $|J_\mathrm{RKKY}^{\alpha\beta}|$ as functions of inter-impurity distance $R$, displayed on a logarithmic scale. For all sublattice configurations, $|J_\mathrm{RKKY}^{\alpha\beta}|$ decay exponentially with $R$. Here the number of unit cells $N=501$ and the hopping ratio $\lambda=2$ for the case without the Kondo coupling to site $C$ (i.e., $x=0$). The dashed lines represent the fits by using the function $J_\mathrm{RKKY}^{\alpha\beta}=c\,e^{-aR}/R^b$, where $a$, $b$, and $c$ are the fitting parameters.
{\Red We obtain $(a,b)=(1.40, 1.32)$, $(1.39, 2.03)$, and $(1.40, 1.33)$ for $J_\mathrm{RKKY}^{AA}$, $J_\mathrm{RKKY}^{AB}$, and $4J_\mathrm{RKKY}^{BB}$, respectively. }
}
\label{fig:JRKKY_lambda2}
\end{figure}

As the case of the stub lattice studied in Ref.~\cite{Laubscher_etal2023}, our numerical results show that the RKKY couplings $J_{\mathrm{RKKY}}^{\alpha\beta}$ decay exponentially with the impurity separation $R$ for all sublattice configurations. Additionally, consistent with the general result for bipartite lattices at half filling~\cite{Saremi2007}, we find that the ground state is FM (AFM) when the two impurities reside on the same (different) sublattices of the bipartition. For illustration, the case of the hopping ratio $\lambda=2$ is displayed in Fig.~\ref{fig:JRKKY_lambda2}. We find that $J_\mathrm{RKKY}^{\alpha\beta}\propto e^{-aR}/R^b$ for all sublattice configurations with identical decay constant $a$. Moreover, the relation $J_\mathrm{RKKY}^{BB}=J_\mathrm{RKKY}^{AA}/4$ is always satisfied in our calculations.

The exponentially decaying behavior is reminiscent of the typical Bloembergen-Rowland behavior found in conventional insulators~\cite{Bloembergen1955,Litvinov2016}. For small $J_{1,2}/t$ (here $J_1/t=J_2/t=0.2$), this can be understood by using second-order perturbation theory, which gives a correction in the ground-state energy
\begin{align}
\Delta E &=\sum_{m\neq0} \frac{
\langle\Psi_0|\tilde{H}_\mathrm{imp}^{(1)}+\tilde{H}_\mathrm{imp}^{(2)}|\Psi_m\rangle
\langle\Psi_m|\tilde{H}_\mathrm{imp}^{(1)}+\tilde{H}_\mathrm{imp}^{(2)}|\Psi_0\rangle
}{E_0 - E_m} \nonumber \\
&\rightarrow \sum_{m\neq0} \frac{
\langle\Psi_0|\tilde{H}_\mathrm{imp}^{(1)}|\Psi_m\rangle
\langle\Psi_m|\tilde{H}_\mathrm{imp}^{(2)}|\Psi_0\rangle
}{E_0 - E_m} + \mathrm{H.c.} \; .
\label{eq:2nd_correction}
\end{align}
Here $|\Psi_0\rangle$ and $|\Psi_m\rangle$ denote the ground state and the excited state with the corresponding energies $E_0$ and $E_m$, respectively. We note that $\Delta E$ needs to be averaged over the subspace of degenerate ground state to achieve the final answer. Besides, only mixed terms in the last step are selected, namely: $\tilde{H}_\mathrm{imp}^{(1)}$ in the first matrix element and $\tilde{H}_\mathrm{imp}^{(2)}$ in the second, and vice versa. This is because the mixed terms take proper account of the mutual influence of the two magnetic impurities. When either magnetic impurity 1 or 2 (or both) is located on the $A$ site, as implied by Eq.~\eqref{eq:detangledA}, only the particle-hole excitations between two dispersive bands matter and the flat-band states play no role. Therefore, the present three-band model behaves as a conventional insulator of two bands, and thus usual Bloembergen-Rowland results will be observed. Interestingly, even if both magnetic impurities locate on the $B$ sites, the flat-band states still have no contribution. This is because $\tilde{H}_\mathrm{imp}^{(1)}$ and $\tilde{H}_\mathrm{imp}^{(2)}$ excite the flat-band states at \emph{different} sites such that the corresponding contributions to numerator of Eq.~\eqref{eq:2nd_correction} become vanishing. Notice that the matrix elements of $V^{(i)}_\sigma$ in Eq.~\eqref{eq:detangledB} is smaller by a factor 2 than those in Eq.\eqref{eq:detangledA}. This explains the observed relation $J_\mathrm{RKKY}^{BB}=J_\mathrm{RKKY}^{AA}/4$.

\begin{figure}[htb]
\includegraphics[width=0.9\linewidth]{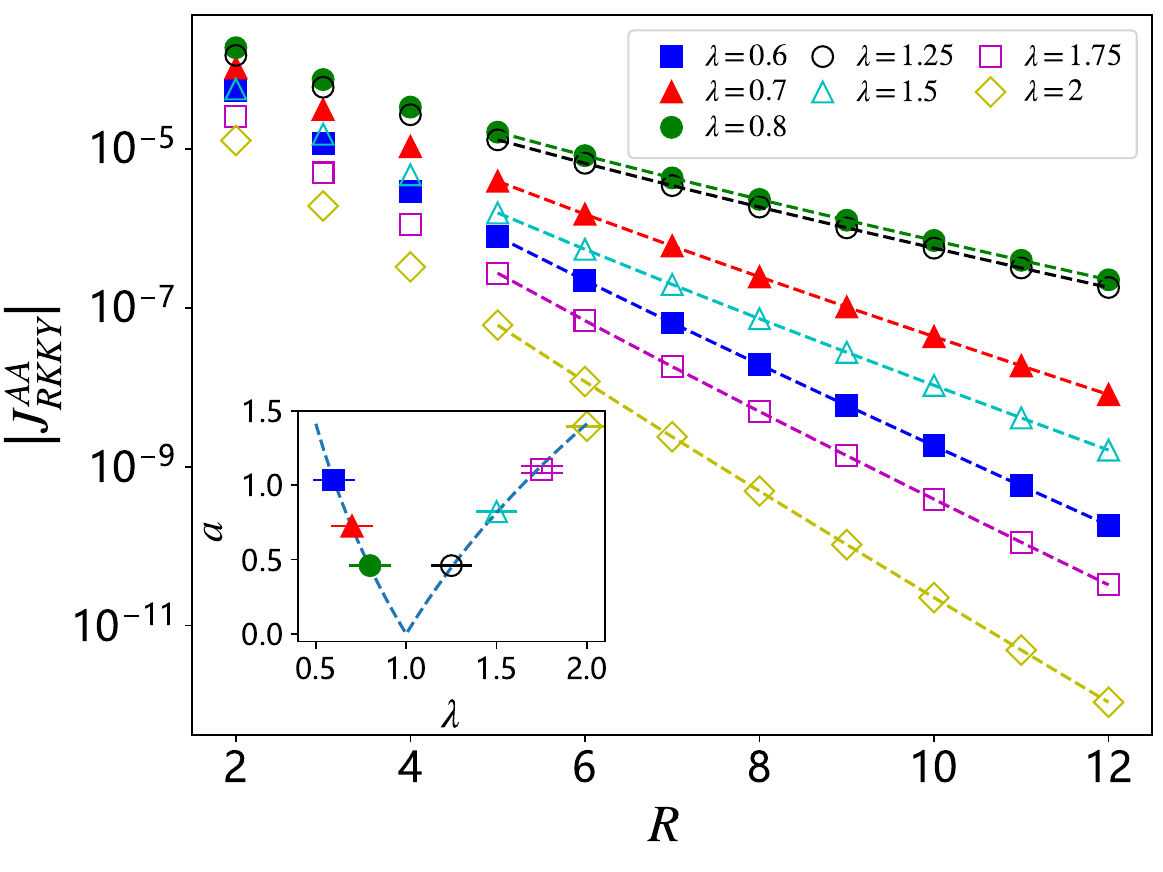}
\caption{Absolute value of the RKKY coupling $|J_\mathrm{RKKY}^{AA}|$ as a function of inter-impurity distance $R$ for various $\lambda$'s and displayed on a logarithmic scale. The dashed lines represent the fits by using the function $J_\mathrm{RKKY}^{\alpha\beta}=c\,e^{-aR}/R^b$, where $a$, $b$, and $c$ are the fitting parameters.
The inset illustrates how the decay constant $a$ varies with $\lambda$.
The dashed lines in the inset represent the analytical result $a=2|1-\lambda|/\sqrt{\lambda}$. Here $N=501$ and $x=0$.
{\Red The values of $b$ exhibit a minor dependence on $\lambda$: $b=1.27$, 1.26, 1.21, 1.21, 1.27, 1.47, and 1.32 for $\lambda=0.6$, 0.7, 0.8, 1.25, 1.5, 1.75, and 2, respectively. }
}
\label{fig:JRKKY_gapful}
\end{figure}

According to the Bloembergen-Rowland theory, the value of the decay constant $a$ is determined by the band gap $E_g$ between two dispersive bands and the effective mass $m^*$ of electrons, i.\,e., $a^{-1}=\hbar/\sqrt{2m^*E_g}$~\cite{Litvinov2016}. If we expand the dispersion relation $\epsilon_+(k)$ of the upper dispersive band around its band bottom at $k=\pi$, we have
\begin{align}
\epsilon_+(k=\pi-\tilde{k})
&\cong\sqrt{2}\,|1-\lambda|+\frac{\lambda}{\sqrt{2}\,|1-\lambda|}\,\tilde{k}^2 \nonumber \\
&\equiv \frac{1}{2}E_g + \frac{\hbar^2}{2m^*}\,\tilde{k}^2 \; .
\end{align}
Therefore, the decay constant is given by
\begin{equation}
a=\frac{2\,|1-\lambda|}{\sqrt{\lambda}} \; .
\label{decay_constant}
\end{equation}
To verify this result, we present the RKKY coupling $J_\mathrm{RKKY}^{AA}$ for various values of $\lambda$ in Fig.~\ref{fig:JRKKY_gapful}. The inset demonstrates the validity of the analytical result.

{\Red %
The preceding discussions should be broadly applicable to systems with an energy gap between the flat and dispersive bands, particularly for weak Kondo couplings $J_{1,2}$. Because the Bloembergen-Rowland theory is based on the second-order perturbation theory, one expect that the dependence of the RKKY couplings on the Kondo coupling strengths will be $J_\mathrm{RKKY}^{\alpha\beta}\propto J_1 J_2$. That is, the general functional form of the RKKY couplings will be $J_\mathrm{RKKY}^{\alpha\beta}\propto J_1 J_2\,e^{-aR}/R^b$, where the decay constant $a$ is given by Eq.~\eqref{decay_constant}. Our data for $J_1/t=J_2/t=0.3$ and 0.4, presented in the Appendix~\ref{sec:App1} (see Fig.~\ref{fig:Jdep_gapful}), provide further support for our picture.
}%

\subsection{Gapless systems with $\lambda=1$}\label{sec:lambda1}

We now investigate the RKKY interaction in gapless systems with $\lambda=1$. Since there is no energy gap, the RKKY couplings $J_{\mathrm{RKKY}}^{\alpha\beta}$ are expected to decay as a power law with respect to the separation distance $R$. The specific exponents could depend on the sublattice configuration. For the case of $x=0$ such that there is no local exchange coupling to the $C$ sublattice, the present model reduces to the case of the diamond lattice discussed in Ref.~\cite{Laubscher_etal2023}. It is found in this particular case that, while $J_\mathrm{RKKY}^{AA}$ follows a conventional one-dimensional metal-like decay of $1/R$, the flat band leads to unusual $1/R^3$ and $1/R^5$ decays for $J_\mathrm{RKKY}^{AB}$ and $J_\mathrm{RKKY}^{BB}$, respectively. These surprising behaviors originate from nonperturbative effects, since the second-order perturbation theory erroneously predicts a $1/R$ decay for \emph{all} sublattice configurations. An intriguing question arises: Do these conclusions remain true when considering generic couplings to the flat band?

\begin{figure}[htb]
\includegraphics[width=0.9\linewidth]{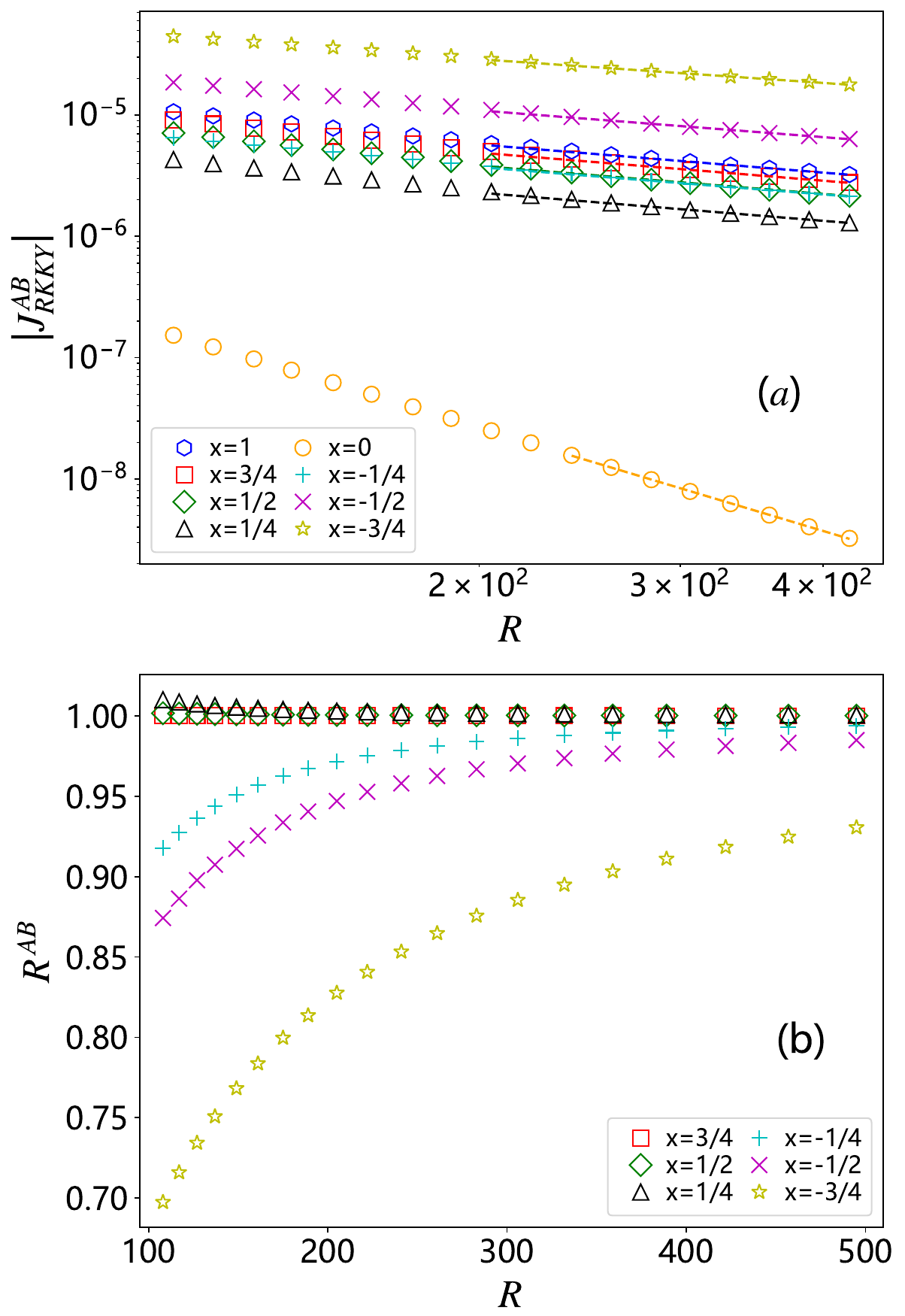}
\caption{(a) Absolute value of the RKKY couplings $|J_\mathrm{RKKY}^{AB}|$ as a function of inter-impurity distance $R$ for various magnetic coupling ratio $x$ and displayed in a log-log scale. Here $\lambda=1$ and the number of unit cells $N=2001$. We note that $J_\mathrm{RKKY}^{AB}$ is positive (negative) when $x$ is positive (negative). The dashed lines represent the fits by using the function $J_\mathrm{RKKY}=c/R^b$, where $b$ and $c$ are the fitting parameters.
(b) The ratio $R^{AB}=[J_\mathrm{RKKY}^{AB}]_x/(z\,[J_\mathrm{RKKY}^{AB}]_{x=1}$) for nonzero $x$'s, where $z=2x/(1+x)$. All the values of $R^{AB}$ tend to unity as $R\rightarrow\infty$.}
\label{fig:JRKKY_gapless1}
\end{figure}
\begin{figure}[htb]
\includegraphics[width=0.9\linewidth]{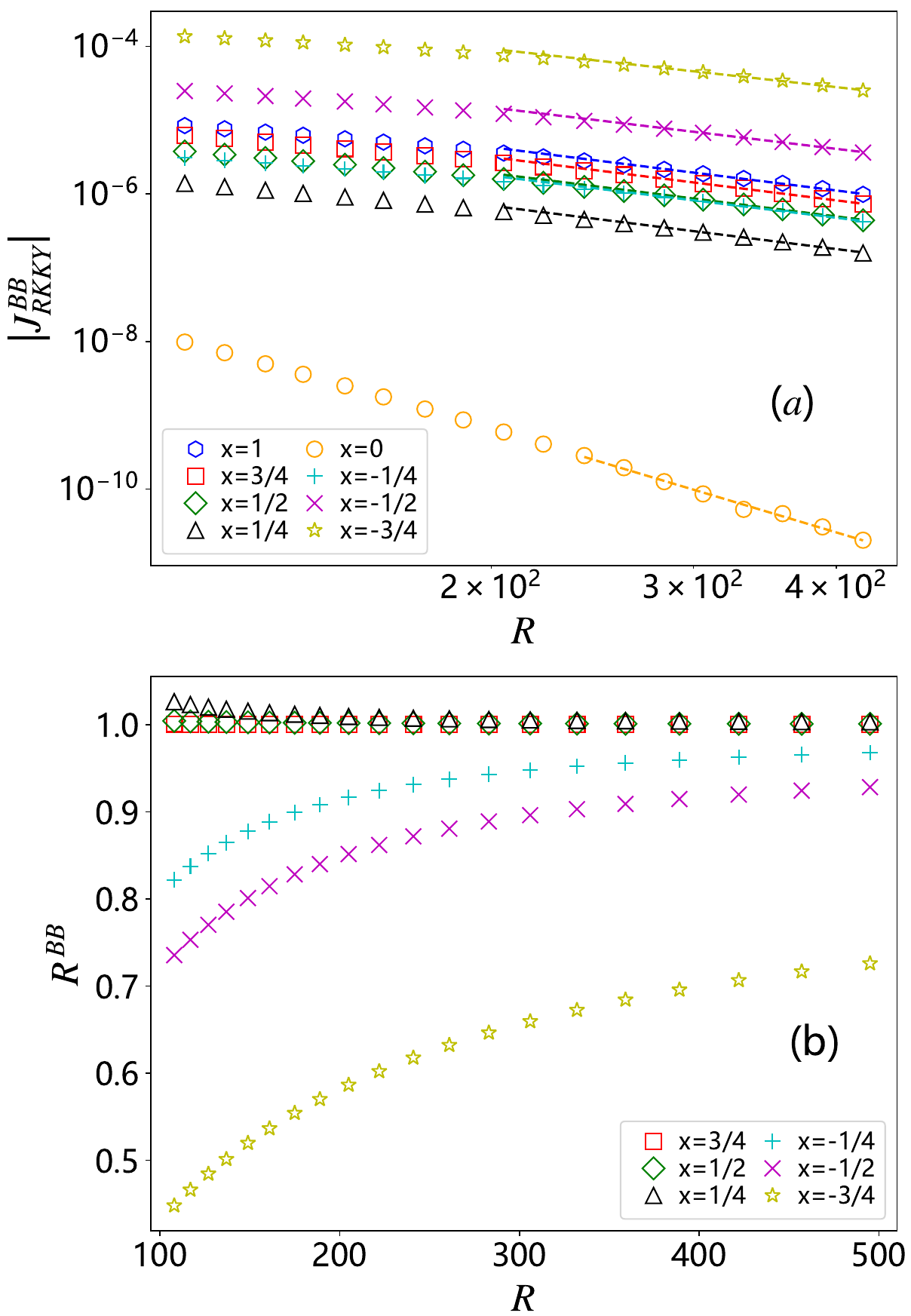}
\caption{(a) Absolute value of the RKKY couplings $|J_\mathrm{RKKY}^{BB}|$ as a function of inter-impurity distance $R$ for various magnetic coupling ratio $x$ and displayed in a log-log scale. Here $\lambda=1$ and the number of unit cells $N=2001$. The dashed lines represent the fits by using the function $J_\mathrm{RKKY}=c/R^b$, where $b$ and $c$ are the fitting parameters.
(b) The ratio $R^{BB}=[J_\mathrm{RKKY}^{BB}]_x/(z^2[ J_\mathrm{RKKY}^{BB}]_{x=1}$) for nonzero $x$'s, where $z=2x/(1+x)$. All the values of $R^{BB}$ tend to unity as $R\rightarrow\infty$.}
\label{fig:JRKKY_gapless2}
\end{figure}

Because the extra coupling to the $C$ sublattice [i.\,e., the second term in Eq.~\eqref{eq:impurtyB}] has no effect on $J_\mathrm{RKKY}^{AA}$, we focus our attention on the behaviors of $J_\mathrm{RKKY}^{AB}$ and $J_\mathrm{RKKY}^{BB}$. Our data obtained by numerically exact diagonalization are presented in Figs.~\ref{fig:JRKKY_gapless1} and \ref{fig:JRKKY_gapless2} for several values of $x$. We find that, in agreement with the general result for bipartite lattices at half filling~\cite{Saremi2007}, $J_\mathrm{RKKY}^{BB}$ always displays FM behavior. However, $J_\mathrm{RKKY}^{AB}$ exhibits AFM behavior for positive $x$'s, but turns to be FM (i.\,e., $J_\mathrm{RKKY}^{AB}<0$) if $x<0$. That is, common wisdom regarding the signs of RKKY interactions breaks down when two impurities are placed on different sublattices and for $x<0$. We note that the breakdown of the theorem proposed in Ref.~\cite{Saremi2007} have been observed in two-dimensional flat-band models~\cite{Oriekhov2020,Bouzerar2021}.
In addition, when $x=-1$ such that the magnetic couplings to the $B$ and the $C$ sublattices have opposite values, $J_\mathrm{RKKY}^{AB}=J_\mathrm{RKKY}^{BB}=0$ up to numerical accuracy. For other $x$'s, as shown in Figs.~\ref{fig:JRKKY_gapless1}\,(a) and \ref{fig:JRKKY_gapless2}\,(a), the RKKY couplings exhibit identical decaying behavior (up to an overall prefactor) for all nonzero values of $x$, while the decay becomes more rapid if $x=0$. That is, the flat band can produce substantial effects such that $J_\mathrm{RKKY}^{AB}\propto1/R^3$ and $J_\mathrm{RKKY}^{BB}\propto1/R^5$ only when $x=0$.

Our findings can be easily explained by using the detangled model in Eqs.~\eqref{eq:detangled0}, \eqref{eq:detangledB}, and~\eqref{eq:detangledA}. To be specific, we set the magnetic impurity 2 on the $B$ site. As discussed in the Appendix~\ref{sec:App2}, our model can be further reduced to a single-chain problem, where the effects of the flat band are incorporated in an effective on-site potential,
\begin{equation}
\tilde{V}^{(2)}_\sigma=V^{(2)}_\sigma + \frac{(t^{(2)}_\sigma)^2}{E-V^{(2)}_\sigma} \; .
\end{equation}
According to Eq.~\eqref{eq:t}, we have $t^{(2)}_\sigma=0$ for $x=1$ and then $\tilde{V}^{(2)}_\sigma=V^{(2)}_\sigma$. The flat-band state thus becomes irrelevant now and conventional result (i.\,e., $J_\mathrm{RKKY}^{AB}$, $J_\mathrm{RKKY}^{BB}\propto1/R$) will be restored.
When $x=-1$, we have instead $V^{(2)}_\sigma=0$ due to Eq.~\eqref{eq:V}. Therefore, $\tilde{V}^{(2)}_\sigma=(t^{(2)}_\sigma)^2/E$, which becomes independent of the orientation of $\mathbf{S_2}$ because of $(S_2)^2=1$. This leads to vanishing $J_\mathrm{RKKY}^{AB}$ and $J_\mathrm{RKKY}^{BB}$ when being evaluated by Eq.~\eqref{eq:RKKY_energy_difference}, which is confirmed by our numerical calculations.

For the case of $x=0$ considered in Ref.~\cite{Laubscher_etal2023}, the relation $V^{(2)}_\sigma=t^{(2)}_\sigma$ is satisfied. The effective on-site potential thus becomes
$\tilde{V}^{(2)}_\sigma=V^{(2)}_\sigma E/(E-V^{(2)}_\sigma)$. From the derivations described in App.~E of Ref.~\cite{Laubscher_etal2023}, we understand that the asymptotic behavior of the RKKY couplings is dominated by electrons around the Fermi surface. Because $\tilde{V}^{(2)}_\sigma$ approaches zero as energy $E$ tends to the Fermi level at $\epsilon_F=0$, this gives significantly weaker RKKY couplings than the conventional ones, as reported in Ref.~\cite{Laubscher_etal2023}.
However, for generic $x$ (but $x\neq0$, $-1$), the effective on-site potential in the vicinity of $E=\epsilon_F=0$ is approximated as a constant,
\begin{equation}
\tilde{V}^{(2)}_\sigma \simeq V^{(2)}_\sigma \left[ 1 - \left(\frac{t^{(2)}_\sigma}{V^{(2)}_\sigma}\right)^2 \right] = z \left[\tilde{V}^{(2)}_\sigma\right]_{x=1} \; ,
\label{effectiveV}
\end{equation}
where $z=2x/(1+x)$. Notice that, just like the case of $x=1$, $\tilde{V}^{(2)}_\sigma$ for $x\neq0$ remains nonzero in the $E\to0$ limit, while it can be either positive or negative depending the sign of $x$. That is, cases other than $x=0$ and $-1$ should have similar asymptotic behavior, as observed in Figs.~\ref{fig:JRKKY_gapless1}\,(a) and \ref{fig:JRKKY_gapless2}\,(a). When compared to the scenario with $x=1$, we anticipate that the asymptotic behavior of $J_\mathrm{RKKY}^{AB}$ and $J_\mathrm{RKKY}^{BB}$ will include additional multiplication factors $z$ and $z^2$, respectively. This is verified by our numerical findings, as shown in Figs.~\ref{fig:JRKKY_gapless1}\,(b) and \ref{fig:JRKKY_gapless2}\,(b).

In summary, for the gapless case of $\lambda=1$, by adjusting the coupling strength with lattice site $C$ ($x\neq0$), the coupling between localized flat-band states and dispersive bands will change accordingly (see the right of Fig.~\ref{fig:dimond_chain}). When $x=0$, the hybridization to the flat-band states causes vanishing effective on-site potential at the Fermi level, and thus makes weaker the asymptotic RKKY interactions ($J_\mathrm{RKKY}^{AB}\sim1/R^3$ and $J_\mathrm{RKKY}^{BB}\sim1/R^5$~\cite{Laubscher_etal2023}). When $x\neq0$, the effective on-site potential at the Fermi level receives a renormalization factor $z=2x/(1+x)$ in comparison with the $x=1$ case. Consequently, the asymptotic RKKY interactions behaves as $J_\mathrm{RKKY}^{AB}\sim z/R$ and $J_\mathrm{RKKY}^{BB}\sim z^2/R$. These conclusions are supported by our numerical findings.

{\Red %
Although we focus on the case of $J_1/t=J_2/t=0.2$, these conclusions should be applicable to generic gapless flat-band systems with weak Kondo couplings $J_{1,2}/t$. Since the asymptotic RKKY interactions for nonzero $x$ behave similarly to those for $x=1$, in which complete detangling of the CLSs (i.\,e., $t^{(i)}_\sigma=0$) occurs and conventional perturbative results become valid, we anticipate  $J_\mathrm{RKKY}^{\alpha\beta}\propto J_1 J_2$ for generic $x$ (excluding $x=0$ and $-1$). Our data for $J_1/t=J_2/t=0.3$ and 0.4, presented in the Appendix~\ref{sec:App1} (see Fig.~\ref{fig:Jdep_gapless}), support this expectation. However, when $x=0$, the hybridization with the flat-band states results in a vanishing $\tilde{V}^{(2)}_\sigma$ at the Fermi level, thereby weakening the asymptotic RKKY interactions and invalidating the perturbative result, as shown in Fig.~\ref{fig:Jdep_gapless} and firstly revealed in Ref.~\cite{Laubscher_etal2023}.
}%

\section{conclusion}\label{sec:conclusion}

In this study, to investigate the RKKY interaction in a flat-band system, we employ a coordinate transformation that detangles flat-band states from dispersive ones. Under the new representation, several features of exact RKKY couplings can be well understood.

For the dimerized diamond lattice under consideration, an immediate consequence is that only the flat-band states directly atop the impurity sites matter. This explains the reason why the RKKY couplings are independent of the precise filling factor. Besides, conventional one-dimensional results will be restored if one sets the ratio $x$ of the Kondo couplings to be unity such that the flat band becomes fully detangled from the dispersive bands. Furthermore, we find that unusual asymptotic behaviors in the RKKY couplings $J_\mathrm{RKKY}^{AB}$ and $J_\mathrm{RKKY}^{BB}$ appear only under particular conditions, that is, $\lambda=1$ and $x=0$. Otherwise, systems will exhibit the typical exponential decay and the $1/R$ decay for the gapped and gapless cases, respectively. However, when $\lambda=1$ and $x<0$, the conventional result for bipartite lattices at half filling~\cite{Saremi2007} regarding the sign of $J_\mathrm{RKKY}^{AB}$ breaks down due to the renormalization effect caused by flat-band states.

The physical picture and the qualitative conclusions of our study would be applicable as well to other one-dimensional flat-band models and even higher dimensional ones. Specifically, the flat band can significantly influence the RKKY interaction only in the gapless case and with specific magnetic coupling configurations.
{\Red %
For instance, the flat band in electronic systems on a Lieb lattice can be isolated from other dispersive bands by energy gaps when intrinsic spin-orbit coupling is activated~\cite{Weeks-Franz2010,MTTran2018}. Our conclusion thus suggests that this spin-orbit coupling could play a crucial role in determining properties of the RKKY interaction.
}%

Our current approach sheds light on the interplay between flat-band states and carrier-mediated exchange interactions, with broader implications for understanding magnetic properties in flat-band materials.
{\Red %
However, this analysis focuses solely on classical spins, justifying the neglect of the Kondo effect. Recently, several studies have investigated Kondo physics in flat-band systems~\cite{MTTran2018,MTTran2019,Kumar_etal2021,Lee2021,Kourris-Vojta2023}.
It would be interesting to explore in future studies the flat-band Kondo physics through the perspective of Fano defects. This could provide deeper insights into the interplay between Kondo screening and the RKKY interaction in flat-band systems.
}%

\begin{acknowledgments}
The authors would like to thank Yu-Wen Lee for enlightening discussions. This research was supported by Grant No. NSTC 112-2112-M-029-005 of the National Science and Technology Council of Taiwan.
\end{acknowledgments}

\appendix
\section{Dependence of RKKY coupling on the Kondo coupling strength}\label{sec:App1}

In the main text, we focus on the case of $J_1/t=J_2/t=0.2$. In this appendix, we illustrate the dependence of the RKKY couplings on the Kondo coupling strengths. For simplicity, we restrict our discussion to the case where $J_1=J_2\equiv J$. We conclude that $J_\mathrm{RKKY}^{\alpha\beta}\propto J^2$ in most scenarios, validating the second-order perturbation theory. However, in the specific case of a gapless spectrum (i.e., hopping ratio $\lambda=1$) and a magnetic coupling ratio of $x=0$, nonperturbative effects become significant.

\begin{figure}[htb]
\includegraphics[width=0.9\linewidth]{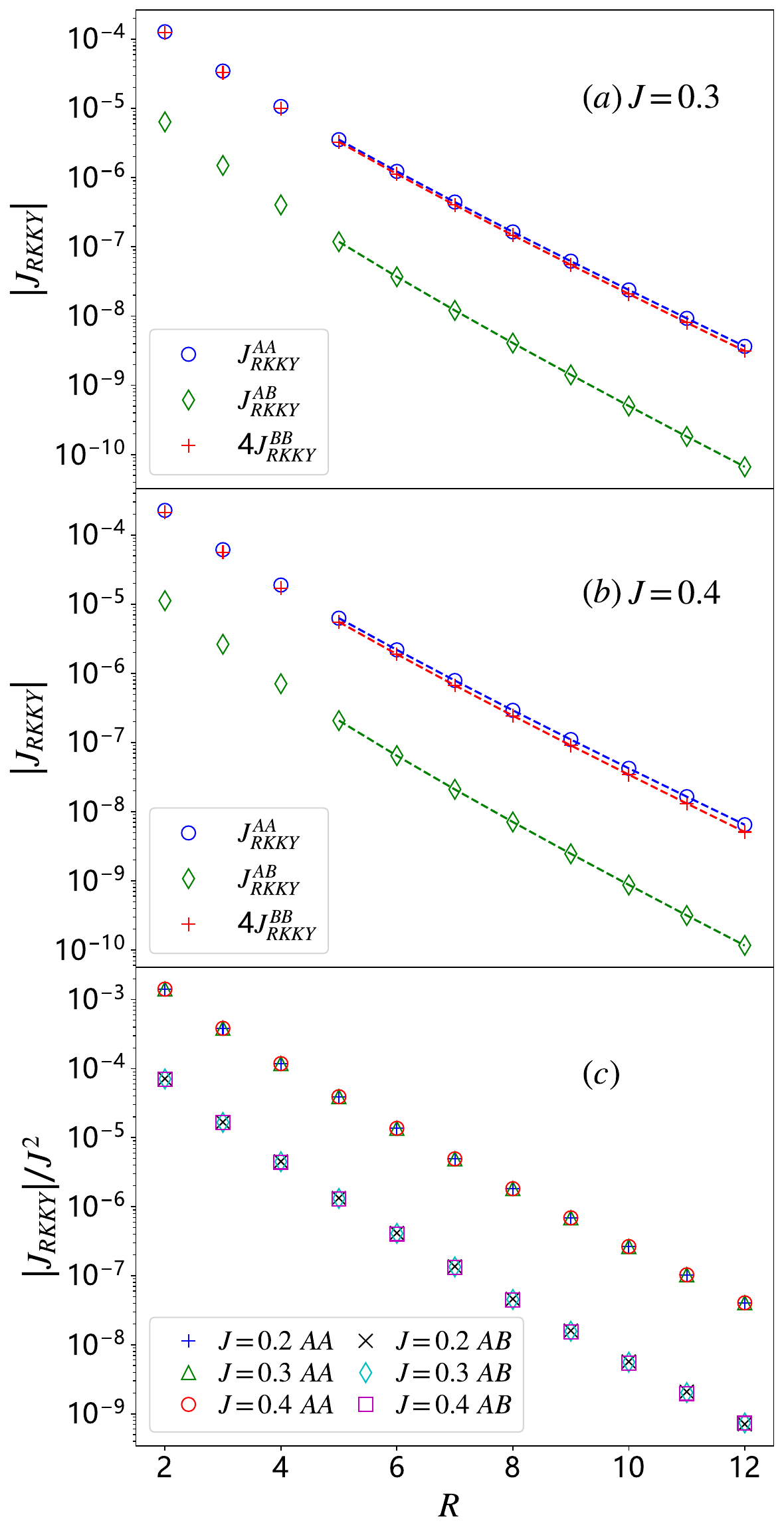}
\caption{%
$|J_\mathrm{RKKY}^{\alpha\beta}|$ as functions of inter-impurity distance $R$, displayed on a logarithmic scale for $J_1/t=J_2/t\equiv J/t=0.3$ (a) and 0.4 (b). Here the number of unit cells $N=501$ and the hopping ratio $\lambda=1.5$ for the case without the Kondo coupling to site $C$ (i.e., $x=0$). For all sublattice configurations, $|J_\mathrm{RKKY}^{\alpha\beta}|$ exhibits exponential decay with $R$. The dashed lines represent the fits by using the function $J_\mathrm{RKKY}^{\alpha\beta}=c\,e^{-aR}/R^b$, where $a$, $b$, and $c$ are the fitting parameters.
(c) Data of $|J_\mathrm{RKKY}^{\alpha\beta}|$ divided by $J^2$ for different $J$ collapse on a universal curve.
}
\label{fig:Jdep_gapful}
\end{figure}

As an example, we consider a gapful system with a hopping ratio of $\lambda=1.5$ and no Kondo coupling to site $C$ (i.e., $x=0$). Our numerical results of $J_\mathrm{RKKY}^{\alpha\beta}$ for $J/t=0.3$ and 0.4 are shown in Fig.~\ref{fig:Jdep_gapful}\,(a) and (b), respectively. The same as the findings presented in Sec.~\ref{sec:lambda2}, $J_\mathrm{RKKY}^{\alpha\beta}$ behave as exponentially decaying functions of inter-impurity distance $R$.
Furthermore, as seen from Fig.~\ref{fig:Jdep_gapful}\,(c), data of $J_\mathrm{RKKY}^{\alpha\beta}$ divided by $J^2$ for different values of $J$ collapse onto a universal curve. This implies that $J_\mathrm{RKKY}^{\alpha\beta}\propto J^2$, in agreement with the Bloembergen-Rowland theory discussed in Sec.~\ref{sec:lambda2}.
We thus conclude that, for systems with an energy gap between the
flat and the dispersive bands, the general form of the RKKY couplings will be $J_\mathrm{RKKY}^{\alpha\beta}\propto J_1 J_2\,e^{-aR}/R^b$, where the decay constant $a$ is given by Eq.~\eqref{decay_constant}.

\begin{figure}[htb]
\includegraphics[width=0.9\linewidth]{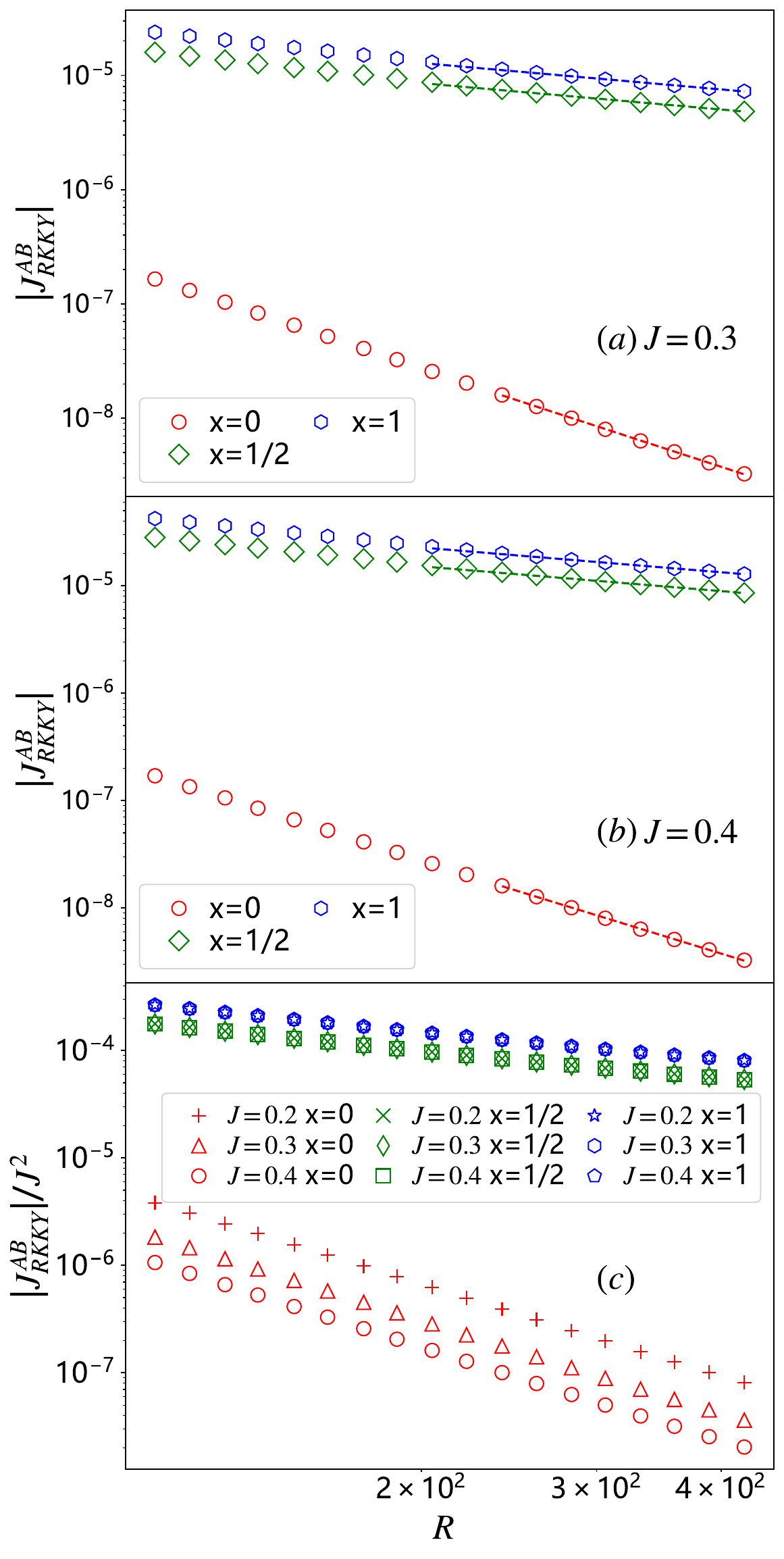}
\caption{%
$|J_\mathrm{RKKY}^{AB}|$ as a function of inter-impurity distance $R$ for three typical magnetic coupling ratio $x$ and for $J_1/t=J_2/t\equiv J/t=0.3$ (a) and 0.4 (b) are displayed in a log-log scale. Here $\lambda=1$ and the number of unit cells $N=2001$. The dashed lines represent the fits by using the function $J_\mathrm{RKKY}=c/R^b$, where $b$ and $c$ are the fitting parameters.
(c) Results of $|J_\mathrm{RKKY}^{AB}|$ divided by $J^2$ for different $J$. When $x\neq0$, our data collapse onto a universal curve, whereas the data collapse fails for $x=0$.
}
\label{fig:Jdep_gapless}
\end{figure}

For the gapless system with $\lambda=1$, we present the numerical results of $J_\mathrm{RKKY}^{AB}$ for illustration. Our findings for $J/t=0.3$ and 0.4 with distinct magnetic coupling ratio $x$ are shown in Fig.~\ref{fig:Jdep_gapless}\,(a) and (b), respectively. The same as the observations found in Sec.~\ref{sec:lambda1}, $J_\mathrm{RKKY}^{AB}$ exhibit identical decaying behavior (up to an overall prefactor) for the cases of $x\neq0$, while the decay becomes more rapid if $x=0$.
Moreover, as seen from Fig.~\ref{fig:Jdep_gapless}\,(c), the data of $J_\mathrm{RKKY}^{AB}$ divided by $J^2$ for different values of $J$ collapse onto a universal curve when $x\neq0$. This implies that $J_\mathrm{RKKY}^{AB}\propto J^2$ in these cases, confirming the validity of the second-order perturbation theory. However, the data collapse does not occur for $x=0$, highlighting the significance of non-perturbative effects as revealed in Ref.~\cite{Laubscher_etal2023}.

As discussed in Sec.~\ref{sec:lambda1}, the nontrivial behaviors of gapless flat-band systems can be understood through the perspective of Fano defects. Just like the $x=1$ case, the effective on-site potential $\tilde{V}^{(2)}_\sigma$ for generic $x$ (excluding $x=0$ and $-1$) always approaches a constant as $E\to0$ [see Eq.~\eqref{effectiveV}]. Therefore, similar asymptotic behavior to that of the $x=1$ case is expected. Since complete detangling of the CLSs (i.\,e., $t^{(i)}_\sigma=0$) occurs at $x=1$, the conventional one-dimensional result based on second-order perturbation theory is anticipated. This explains the reason why $J_\mathrm{RKKY}^{AB}\propto J^2$ for nonzero $x$ observed in Fig.~\ref{fig:Jdep_gapless}\,(c). However, when $x=0$, the hybridization to the flat-band states leads to a vanishing effective on-site potential $\tilde{V}^{(2)}_\sigma$ at the Fermi level, thereby weakening the asymptotic RKKY interactions and invalidating the perturbative result.

\section{on-site potential of effective single-chain model}\label{sec:App2}

The local coordinate transformation in Eq.~\eqref{eq:fano-DC} yields a dispersive coordinate $d_{n,A,\sigma}$, $d_{n,B,\sigma}$ and a flat-band coordinate $f_{n,\sigma}$. This results in a generalized Fano-Anderson chain, as shown in the right of Fig.~\ref{fig:dimond_chain}.

Following the approach discussed in Refs.~\cite{Flach_etal2014,Flach_etal2014,Bodyfelt_etal2014,Danieli_etal2015}, by eliminating the flat-band variables, the Fano-Anderson chain can be further reduced to a single-chain problem with a correction in on-site potential.
To be specific, we consider the magnetic impurity 2 being located on the $B$ site of the $n$-th unit cell and set $\lambda=1$. The eigenvalue problem of the detangled model in Eqs.~\eqref{eq:detangled0}-\eqref{eq:detangledB} for the variables at this site reads
\begin{align}
E\,d_{n,B,\sigma}&= V^{(2)}_\sigma\,d_{n,B,\sigma} + t^{(2)}_\sigma f_{n,\sigma} \nonumber \\
&\quad + \sqrt{2}\left( d_{n,A,\sigma} + d_{n+1,A,\sigma} \right) \; , \\
E\,f_{n,\sigma}&= V^{(2)}_\sigma f_{n,\sigma} + t^{(2)}_\sigma\,d_{n,B,\sigma} \; .
\end{align}
Expressing the flat-band variable $f_{n,\sigma}$ through $d_{n,B,\sigma}$, we reduce these equations to a tight-binding form which contains the dispersive portion only:
\begin{equation}
E\,d_{n,B,\sigma}= \tilde{V}^{(2)}_\sigma\,d_{n,B,\sigma} + \sqrt{2}\left( d_{n,A,\sigma} + d_{n+1,A,\sigma} \right) \; .
\end{equation}
Here $\tilde{V}^{(2)}_\sigma$ denotes an effective on-site potential for $d_{n,B,\sigma}$, which is given by
\begin{equation}
\tilde{V}^{(2)}_\sigma=V^{(2)}_\sigma + \frac{(t^{(2)}_\sigma)^2}{E-V^{(2)}_\sigma} \; .
\end{equation}



\end{document}